\documentclass[amsmath,amssymb,aps,twocolumn,nofootinbib,preprintnumbers]{revtex4-2}
\pdfoutput=1

\usepackage{amsmath}
\usepackage{booktabs}
\usepackage{hyperref}
\usepackage{graphicx}
\usepackage{multirow}
\usepackage{siunitx}
\usepackage{xcolor}

\hypersetup{
  colorlinks=true,
  linkcolor=blue,
  citecolor=blue,
  urlcolor=blue
}

\begin{document}

\preprint{DESY-22-126}
\preprint{IPPP/22/51}

\title{QFitter -- A Quantum Fitting Framework Applied to Effective Field Theories}

\author{Juan Carlos Criado}
\email{juan.c.criado@durham.ac.uk}
\affiliation{Institute for Particle Physics Phenomenology, Durham University, Durham DH1 3LE, UK}
\affiliation{Department of Physics,  Durham University, Durham DH1 3LE, UK}

\author{Roman Kogler}
\email{roman.kogler@desy.de}
\affiliation{Deutsches Elektronen-Synchrotron DESY, Notkestr.\ 85, 22607 Hamburg, Germany}

\author{Michael Spannowsky}
\email{michael.spannowsky@durham.ac.uk}
\affiliation{Institute for Particle Physics Phenomenology, Durham University, Durham DH1 3LE, UK}
\affiliation{Department of Physics, Durham University, Durham DH1 3LE, UK}

\begin{abstract}
The use of experimental data to constrain the values of the Wilson coefficients of an Effective Field Theory (EFT) involves minimising a $\chi^2$ function that may contain local minima. Classical optimisation algorithms can become trapped in these minima, preventing the determination of the global minimum. The quantum annealing framework has the potential to overcome this limitation and reliably find the global minimum of non-convex functions. We present QFitter, a quantum annealing method to perform EFT fits. Using a state-of-the-art quantum annealer, we show with concrete examples that QFitter can be used to fit sets of at least eight coefficients, including their quadratic contributions. An arbitrary number of observables can be included without changing the required number of qubits. We provide an example in which $\chi^2$ is non-convex and show that QFitter can find the global minimum more accurately than its classical alternatives.
\end{abstract}

\maketitle

\section{Introduction}

Determining model parameters by adjusting a theoretical hypothesis until it best fits experimental data is commonly referred to as \emph{fitting}. 
Fitting algorithms use optimisation methods which aim to find the model's parameters that result in the smallest value of a characteristic function, like the negative log-likelihood function, $\chi^2$, which incorporates all experimental observables of interest simultaneously. As such procedure allows an interpretation of experimental data in terms of a given theoretical model, fitting algorithms are at the core of phenomenological studies of new physics.

Effective Field Theories (EFTs) provide a model-independent framework to tension experimental data with the imprints of new physics scenarios, assuming that any new particles are significantly heavier than the Standard Model particles. The free parameters of an EFT are the coefficients of the local operators that appear in its Lagrangian. They are known as the \emph{Wilson coefficients}. The EFT formulation ensures that the Wilson coefficients can capture any effects of the heavy particles at low energies compared to their masses.

Because of this, EFT fits have become an essential tool in interpreting experimental data in the current context, in which no resonant new physics has been observed at the LHC. Subsets of the coefficients of the EFT for the non-linear realisation of the electroweak symmetry have been fitted to the relevant experimental data in Refs.~\cite{Buchalla:2015qju, Brivio:2016fzo}. Similarly, fits for the linear realisation have been performed in Refs.~\cite{Englert:2015hrx, Buckley:2015lku, Klute:2012pu, Englert:2017aqb, Ellis:2018gqa, Ethier:2021bye, deBlas:2022ofj,Anisha:2021hgc}, for example.

If non-linear contributions of the Wilson coefficients are considered in the fit, the $\chi^2$ function to be minimised may develop local minima which differ from the global one. Furthermore, local minima can also occur when some nuisance parameters are included. This may prevent classical optimisation algorithms from reaching the values of the Wilson coefficients that fit the observed experimental values optimally. Indeed, such algorithms usually obtain the solution by incrementally improving a given point in parameter space, which can lead to them becoming trapped in a local minimum.

Quantum annealing provides an optimisation framework with the potential to perform better than classical algorithms in minimising non-convex functions~\cite{farhi02a, Abel:2020ebj, Abel:2020qzm,Abel:2021fpn, Abel:2022lqr}. The availability of physical quantum annealing devices with thousands of qubits has made it possible to apply this approach to real-world problems in recent years~\cite{finilla94a, kadowaki98a, brooke99a, dickson13a, lanting14a, albash15a, albash16a, boixo16a, chancellor16b, Benedetti16a, Muthukrishnan2016, cervera, LantingAQC2017}.

In this work, we construct a formulation of EFT fitting calculations that can be implemented in these devices. An arbitrary number of observables can be included in the fit without changing the required number of qubits. Furthermore, general polynomial dependence of the theoretical predictions on the Wilson coefficients can be implemented in this formulation, with non-linear terms encoded by auxiliary qubits. In practice, we find that current quantum annealing devices can be used to perform fits with at least 8 Wilson coefficients and quadratic dependence of observables on them. Using practical examples, we show that the quantum approach generates the best-fit estimator for the coefficients accurately and more consistently than classical methods in a non-convex problem.

The rest of this paper is organised as follows. In Section~\ref{sec:method}, we present the QFitter method for fitting EFT coefficients using quantum annealing. In Section~\ref{sec:results}, we provide the results of applying QFitter to three different problems, including a fit in which the $\chi^2$ function to be minimised is non-convex. We compare QFitter to several classical approaches. We give our conclusions in Section~\ref{sec:conclusions}.

\section{Method}
\label{sec:method}

\subsection{Quantum annealing}
\label{sec:qa}
Quantum annealing is a method for finding the ground state of a given Hamiltonian $H_1$. The Hamiltonian of a quantum annealing device takes the form
\begin{equation}
 H = A(s) H_0 + B(s) H_1,
\end{equation}
where $s$ is a free parameter that can be controlled externally; $A(s)$ and $B(s)$ are continuous functions such that $A(0) > 0 = B(0)$ and $A(1) = 0 < B(1)$; and $H_0$ is a Hamiltonian whose ground state is known in advance.

The solution to the problem is obtained by preparing the system in the ground state of $H_0$ and changing $s$ continuously from $s = 0$ at an initial time $t_i$ to $s= 1$ at a final time $t_f$. The function $s(t)$ described by the time evolution of $s$ is known as the \emph{schedule}. The adiabatic theorem ensures that, if the change in $s$ is sufficiently slow, the system is likely to end in the ground state of the target Hamiltonian $H_1$.

A concrete realization of this method is transverse-field quantum annealing, which has been implemented in real-world devices. In this realization, the system can be viewed effectively as a collection of qubits (that is, quantum systems with two independent states), with the Hamiltonians $H_0$ and $H_1$ given by
\begin{equation}
 H_0 = \sum_i \hat{\sigma}^i_x, \quad
 H_1 = \sum_i h_i \hat{\sigma}^i_z
 + \sum_{ij} J_{ij} \, \hat{\sigma}^i_z \otimes \hat{\sigma}^i_z.
\end{equation}
where $h_i$ and $J_{ij}$ are adjustable parameters, and $\hat{\sigma}^i_{x,z}$ are the $x,z$ Pauli matrices acting on the $i$th qubit. To perform a computation using transverse-field quantum annealing, one needs to encode its result as the ground state of the Ising Hamiltonian $H_1$.

The D-Wave devices provide a physical implementation of this setup, in which not all $J_{ij}$ couplings can be set to a non-vanishing value. In the state-of-the-art \texttt{Advantage\_system} architecture, there are more than 5000 available qubits, but each one is coupled only to 15 others. To find the ground state of Ising models with a higher degree of connectivity, several qubits are chained together with large coupling to act as a single qubit with more connections. The mapping between the abstract Ising model Hamiltonian to be minimized and the one implemented in the physical device is known as an \emph{embedding}.

Once an embedding has been found, and the schedule $s(t)$ and $h_i$, $J_{ij}$ parameters are set, the annealer is typically run several times to reduce the effects of external noise. Then, the final state with the least energy obtained from the different runs is selected. The number of runs is referred to in the context as the \emph{number of reads}.

\subsection{QUBO formulation}

The eigenstates of the quantum Ising Hamiltonian $H_1$ correspond to the states of its classical analogue, whose Hamiltonian is
\begin{equation}
 H_{\text{classical}} = \sum_i h_i \sigma_i + \sum_{ij} J_{ij} \sigma_i \sigma_j,
\end{equation}
with the $\sigma_i$ being classical variables taking the values $\sigma_i = \pm 1$. Thus, the problem solved by the transverse-field quantum annealers can be viewed equivalently as finding the set of values for the $\sigma_i$ such that $H_{\text{classical}}$ is minimized:

This problem can be solved both using quantum annealing and classical algorithms, such as simulated annealing. Quantum annealing has been shown to be more consistent in finding the ground state of some non-convex functions~\cite{Abel:2021fpn}. In Section~\ref{sec:results}, we will compare the performance of quantum annealing to several classical methods for EFT fits.

A useful reformulation of the classical Ising Hamiltonian minimization problem is obtained by making use of the binary variables $\tau_i = (\sigma_i + 1) / 2$, whose possible values are 0 or 1. In terms of them, the problem can be expressed as the minimization of a homogeneous quadratic polynomial:
\begin{equation}
 \min_{\tau_i = 0, 1} \tau_i Q_{ij} \tau_j.
\end{equation}
This is known as a Quadratic Unconstrained Binary Optimization (QUBO) problem. We will refer to the function $\mathcal{L} = \tau_i Q_{ij} \tau_j$ to be minimized in such a problem as the \emph{loss function}.

\subsection{Log-likelihood as a QUBO}

We now tackle the task of finding a QUBO formulation for fits of EFT Wilson coefficients to observables. Let $O^{(\text{exp})}_i$ be the experimentally-measured values of the observables under consideration, and $O^{(\text{th})}_i(c)$ the corresponding theoretical predictions, as functions of the collection of Wilson coefficients $c = (c_1, \ldots, c_M)$. We assume a Gaussian likelihood $L \propto e^{-\chi^2 / 2}$, with
\begin{equation}
  \chi^2 = \sum_{ij} V_a C^{-1}_{ab} V_b, \qquad
  V_a = O^{(\text{exp})}_a - O^{(\text{th})}_a(c),
  \label{eq:chi2}
\end{equation}
where $C^{-1}$ is the inverse covariance matrix. The maximum-likelihood estimator for the coefficients can be obtained by minimizing $\chi^2$.

In any EFT, the theoretical predictions $O^{(\text{th})}_a(c)$ are computed as a series in inverse powers of the cutoff scale $\Lambda$, which is cut at a fixed power, depending on the target precision of the calculation. Each Wilson coefficient is associated with an inverse power of $\Lambda$. The $O^{(\text{th})}_a(c)$ will thus be polynomials in $c$. For simplicity, we restrict ourselves here to quadratic polynomials
\begin{equation}
  O^{(\text{th})}_a(c) = A_a + \sum_i B_{ai} c_i + \sum_{ij} C_{aij} c_i c_j,
  \label{eq:Oth}
\end{equation}
although the method that we present can be straightforwardly extended to higher degrees. Nevertheless, quadratic polynomials are sufficient for most current EFT applications. In particular, dimension-8 and dimension-6 squared contributions in the Standard Model EFT can be dealt with in this setting.

We thus have a $\chi^2$ function to be minimized that is a quartic polynomial in real variables $c_i$. To turn this into a QUBO, we need to re-formulate it as a quadratic polynomial in binary variables. Binarization can be achieved by means of a binary encoding~\cite{Abel:2022lqr, Criado:2022deo}:
\begin{equation}
  c_i = L_i + \frac{U_i - L_i}{1 - 2^{-n-1}} \sum_{n = 1}^N~ \frac{c^{(n)}_i}{2^{n+1}},
  \label{eq:binary-enc}
\end{equation}
with $L_i$, $U_i$ and $N$ being fixed parameters, and the $c^{(n)}_i$ taking binary values 0 or 1. The coefficients $c_i$ can take $2^N$ different values, uniformly distributed in the $[L_i, U_i]$ interval. Substituting this in Eq.~\eqref{eq:Oth} turns $O^{(\text{th})}_a(c)$ into a quadratic function of the binary variables, which means that $\chi^2$ is a quartic polynomial in them.

The reduction from a fourth-degree polynomial to a second-degree one can be made by means of auxiliary binary variables $c^{(m,n)}_{ij}$, representing the products $c^{(m)}_i c^{(n)}_j$. In this way, the $c_i c_j$ factor that appears in the last term of Eq.~\eqref{eq:Oth}, can be written as a linear function in the binary variables
\begin{multline}
  c_i c_j = L_i L_j + \left[
	L_j  \frac{U_i - L_i}{1 - 2^{-n-1}}
	\sum_{n = 1}^N~ \frac{c^{(n)}_i}{2^{n+1}} + (i \leftrightarrow j)
  \right]
  \\
  + \frac{(U_i - L_i)(U_j - L_j)}{(1 - 2^{-n-1})^2}
  \sum_{m,n=1}^N \frac{c^{(m,n)}_{ij}}{2 ^{m+n+2}}.
  \label{eq:binary-enc-quad}
\end{multline}
Replacing Eq.~\eqref{eq:binary-enc} in the linear term of Eq.~\eqref{eq:Oth}, and Eq.~\eqref{eq:binary-enc-quad} in the quadratic one, $O^{(\text{th})}_a(c)$ and $\chi^2$ become linear and quadratic in the binary variables, respectively.

In order for this procedure to work, the constraints $c^{(m,n)}_{ij} = c^{(m)}_i c^{(n)}_j$ need to be enforced somehow. This can be done by means of a \emph{constraint Hamiltonian}: a quadratic function $P(x, y, z)$ of binary variables $x$, $y$, $z$ that achieves its minimum $P = 0$ if and only if $xy = z$. In particular, we use the function
\begin{equation}
  P(x, y, z) = xy - 2z(x + y) + 3z.
\end{equation}
Now, to construct the loss function, we add together $\chi^2$, viewed as a quadratic function of the binary variables, as described above, and the constraint Hamiltonians for all the relations between them:
\begin{equation}
  \mathcal{L} =
  \chi^2
  + \lambda \sum_{ijmn} P\left(c^{(m)}_i, c^{(n)}_j, c^{(m,n)}_{ij}\right).
\end{equation}
When the coefficient $\lambda$ is large enough, the minimum of $\mathcal{L}$ will correspond to the minimum of $\chi^2$ over the set of values of the binary variables that satisfy the constraints. This concludes the re-formulation of the problem as QUBO: the maximum likelihood estimator for the Wilson coefficients can be obtained by minimizing a function quadratic function of binary variables $\mathcal{L}$.

The QUBO problem we have obtained can be solved by the usual methods, including quantum and simulated annealing. We will show in Section~\ref{sec:results} that, in practice, quantum annealing is the most effective of the two for this purpose, especially when $\chi^2$ is a non-convex function of the coefficients $c_i$. The result of any of these methods will be a set of values of the binary variables $c^{(n)}_i$ and $c^{(m,n)}_{ij}$ representing the ground state of $\mathcal{L}$. The fitted values of the $c_i$ can be recovered using Eqs.~\eqref{eq:binary-enc} and~\eqref{eq:binary-enc-quad}.

The number $n_{\text{bin}}$ of binary variables employed in this formulation is highly relevant for the practical applications using quantum annealing devices since, together with the connectivity of the QUBO, it will control the number of qubits in the final embedding. $n_{\text{bin}}$ is controlled by the number $M$ of Wilson coefficients involved in the fit and the number $N$ of binary variables used in the encoding for each of them. In general, one has
\begin{equation}
  n_{\text{bin}} \leq N M + \frac{1}{2} (N + 1) N M^2.
  \label{eq:n-qubits}
\end{equation}
The first term is the number of binary variables required to encode the coefficients. The second one is for their products: there are $N (N + 1) / 2$ different combinations  of the form $c_i c_j$, and at most $M^2$ different choices for $m$ and $n$ given $(i, j)$. In many situations, only certain combinations of coefficients appear in the quadratic terms in Eq.~\eqref{eq:Oth}. In the examples of Section~\ref{sec:results}, only the squares of some individual coefficients appear. Then, the number of products of binary variables to encode is $N M (M - 1) / 2$, and
\begin{equation}
  n_{\text{bin}} \leq N M (M + 1) / 2.
\end{equation}
Concerning the connections, an entry of the $Q$ matrix of the QUBO is non-vanishing if and only if there is one observable to which the corresponding coefficients (or product of coefficients) contribute simultaneously. Since there are observables to which many coefficients may contribute, the degree of connectivity is typically high.

Finally, notice here that the number of observables does not play any role in determining $n_{\text{bin}}$. That means that the number of observables that can be included in the fit is not limited by the available number of qubits when the QUBO is embedded in physical annealing devices.

\subsection{Zooming}

As discussed in Section~\ref{sec:qa}, the total number of available qubits in state-of-the-art devices is in the order of several thousand. However, for problems with a high degree of connectivity that require embedding several physical qubits per abstract qubit in the original problem, the effective number of available qubits is much lower, reaching around 100 for fully connected QUBOs. In an EFT fit, the number $M$ of coefficients is fixed, so the only way to reduce the number of qubits, according to Eq.~\eqref{eq:n-qubits} is to reduce the number $N$ of binary variables per coefficient. Thus, the bound in the number of qubits translates into a bound in $N$, which limits the precision that can be achieved for the best-fit values of the coefficients.

We overcome this limitation using a zooming process, similar to the one presented in Ref.~\cite{Zlokapa:2019lvv}. The idea is to perform the fit in several steps, which we call \emph{epochs}. Each epoch consists of a quantum annealing run for a refined version of the QUBO problem of the previous run, in which the range $[L_i, U_i]$ for each coefficient $c_i$ is updated. Given a parameter $0 < f < 1$, the \emph{zoom factor}, the update rule for the lower and upper limits of the range is
\begin{equation}
 L_i \to c_i - \frac{f}{2} (U_i - L_i),
 \quad
 U_i \to c_i + \frac{f}{2} (U_i - L_i),
\end{equation}
where $c_i$ denotes the value of the corresponding coefficient obtained from the previous epoch. The effect of this transformation is to reduce the length of the $[L_i, U_i]$ range by a factor $f$, while centering it around $c_i$.

The zooming process thus allows achieving any desired precision at the price of introducing a classical update step between quantum annealing runs. We stress that its use is necessary for embedding in current quantum annealing devices because of the limited amount of available qubits. In future devices with a larger quantity of qubits, it might be possible to use a larger number $N$ of binary variables per coefficient, thus making it possible to reach the relevant precision in a single quantum annealing run.

\section{Results and discussion}
\label{sec:results}

\subsection{EWPO fit}
\label{sec:ewpo-fit}

\begin{figure}
  \centering
  \includegraphics[width=0.45\textwidth]{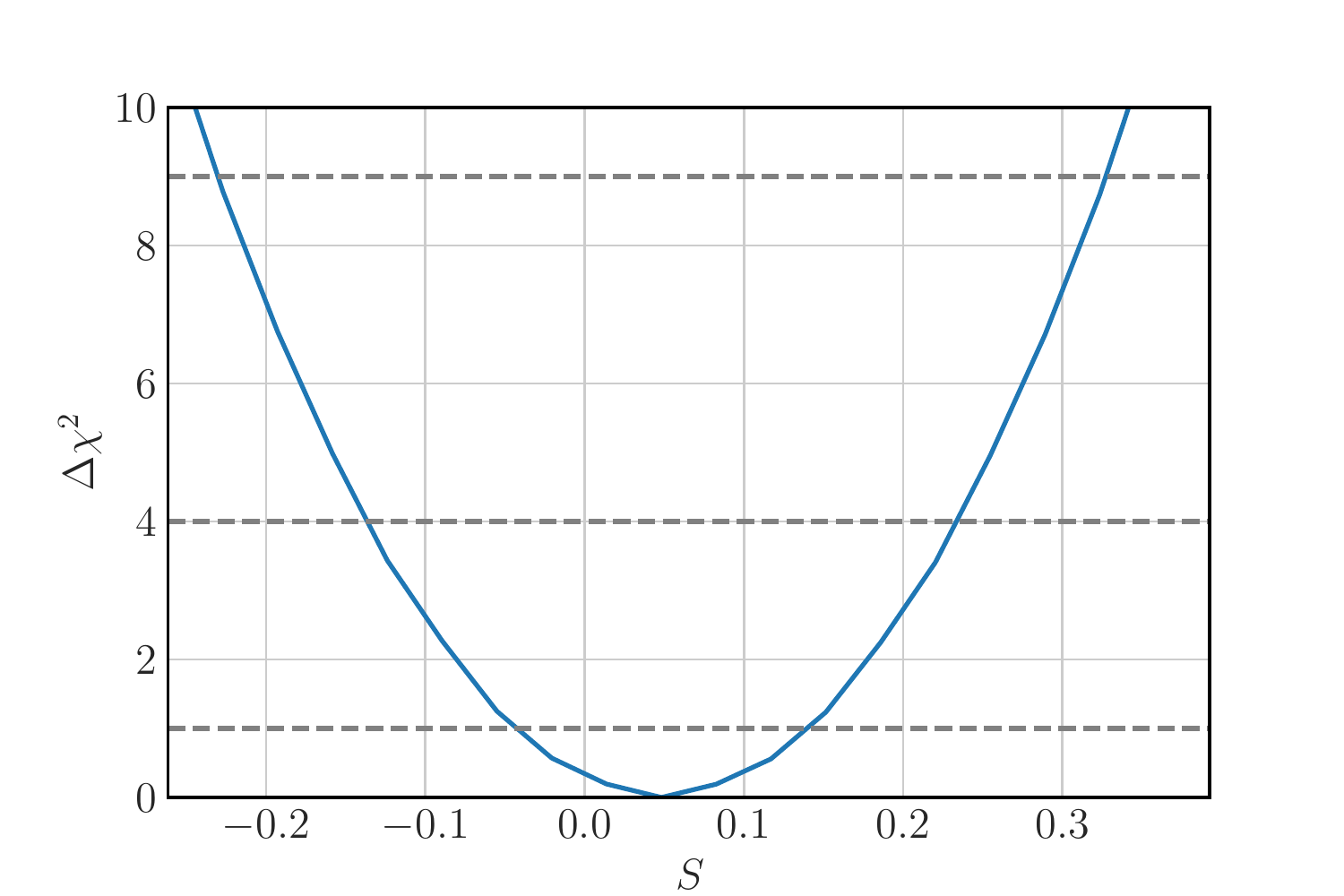}
  \includegraphics[width=0.45\textwidth]{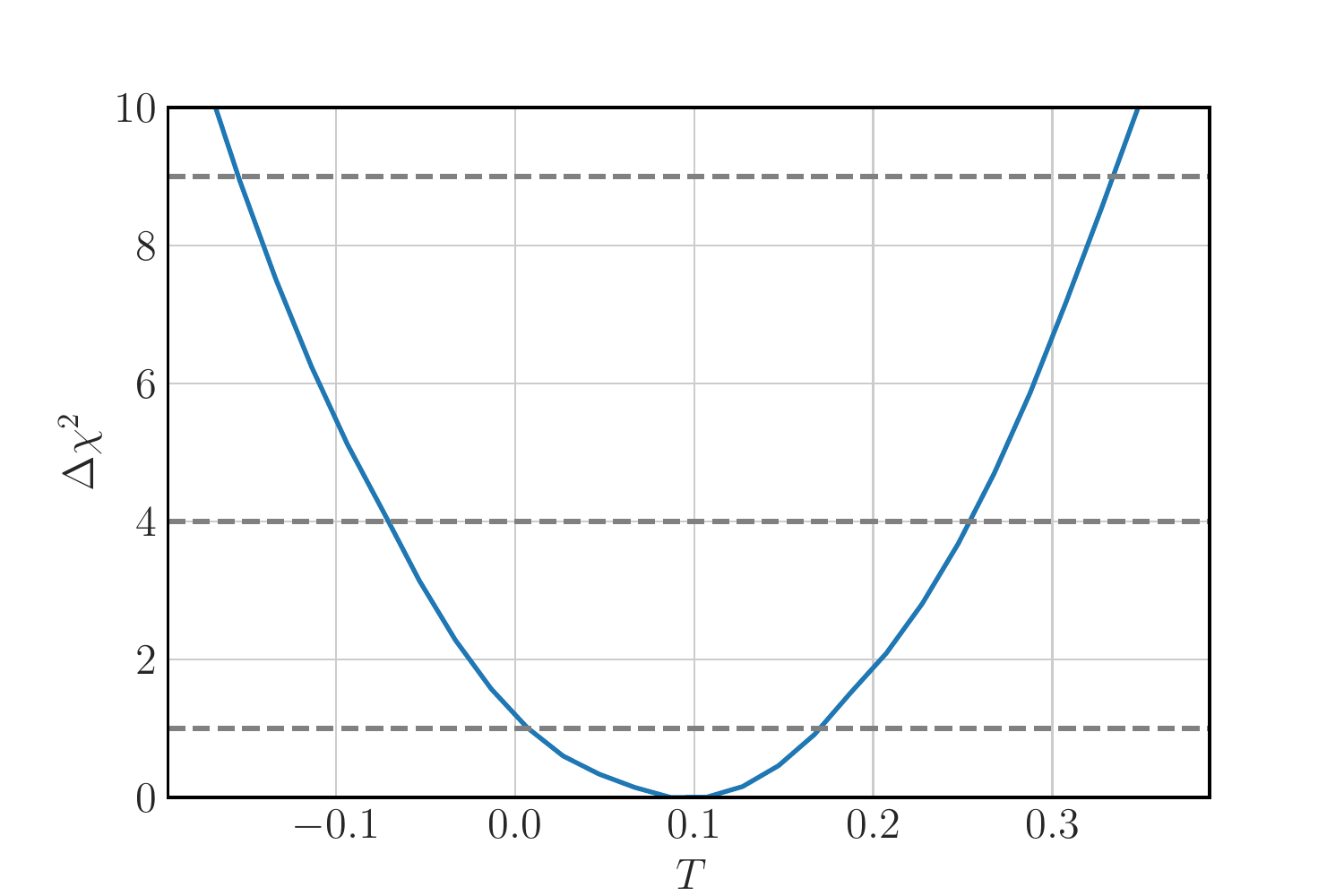}
  \caption{$\Delta \chi^2 = \chi^2 - \chi^2_{\text{min}}$ profiles for the EWPO fit. For each point in the graphs, $\chi^2$ is minimized while the parameter in the $x$ axis is kept fixed.}
  \label{fig:ewpo-profiles}
\end{figure}

\begin{figure*}
  \centering
  \includegraphics[width=0.45\textwidth]{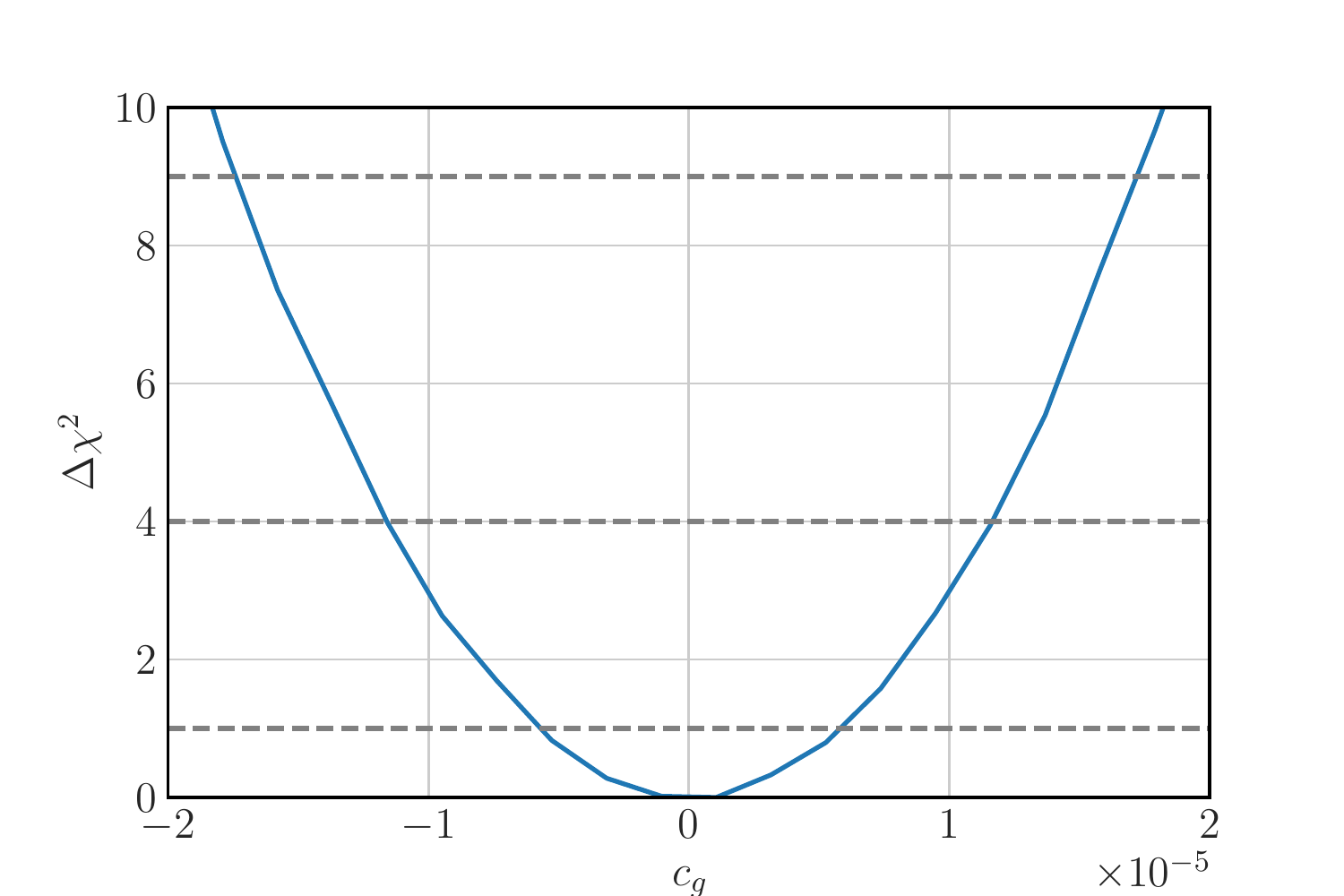}
  \includegraphics[width=0.45\textwidth]{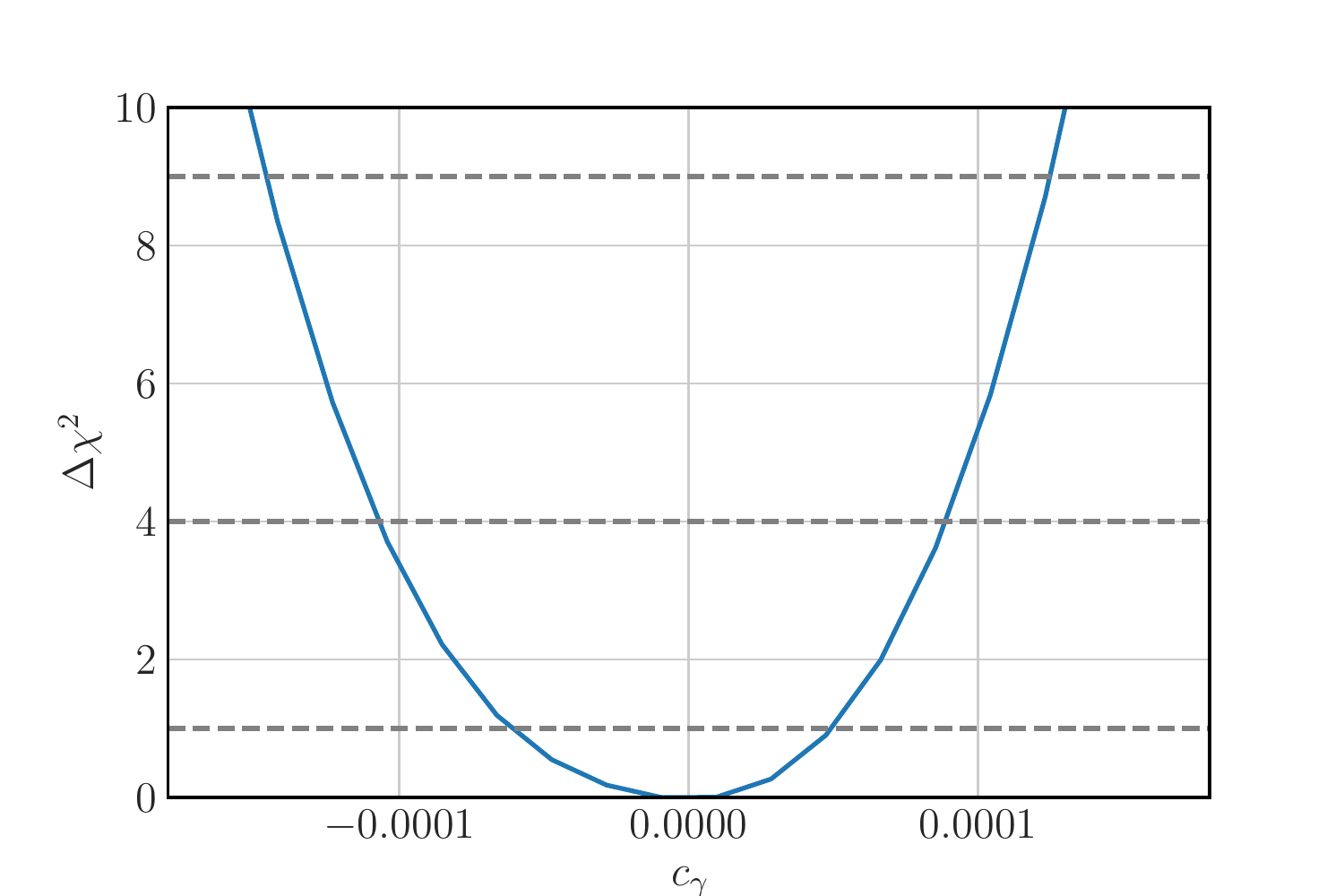}
  \includegraphics[width=0.45\textwidth]{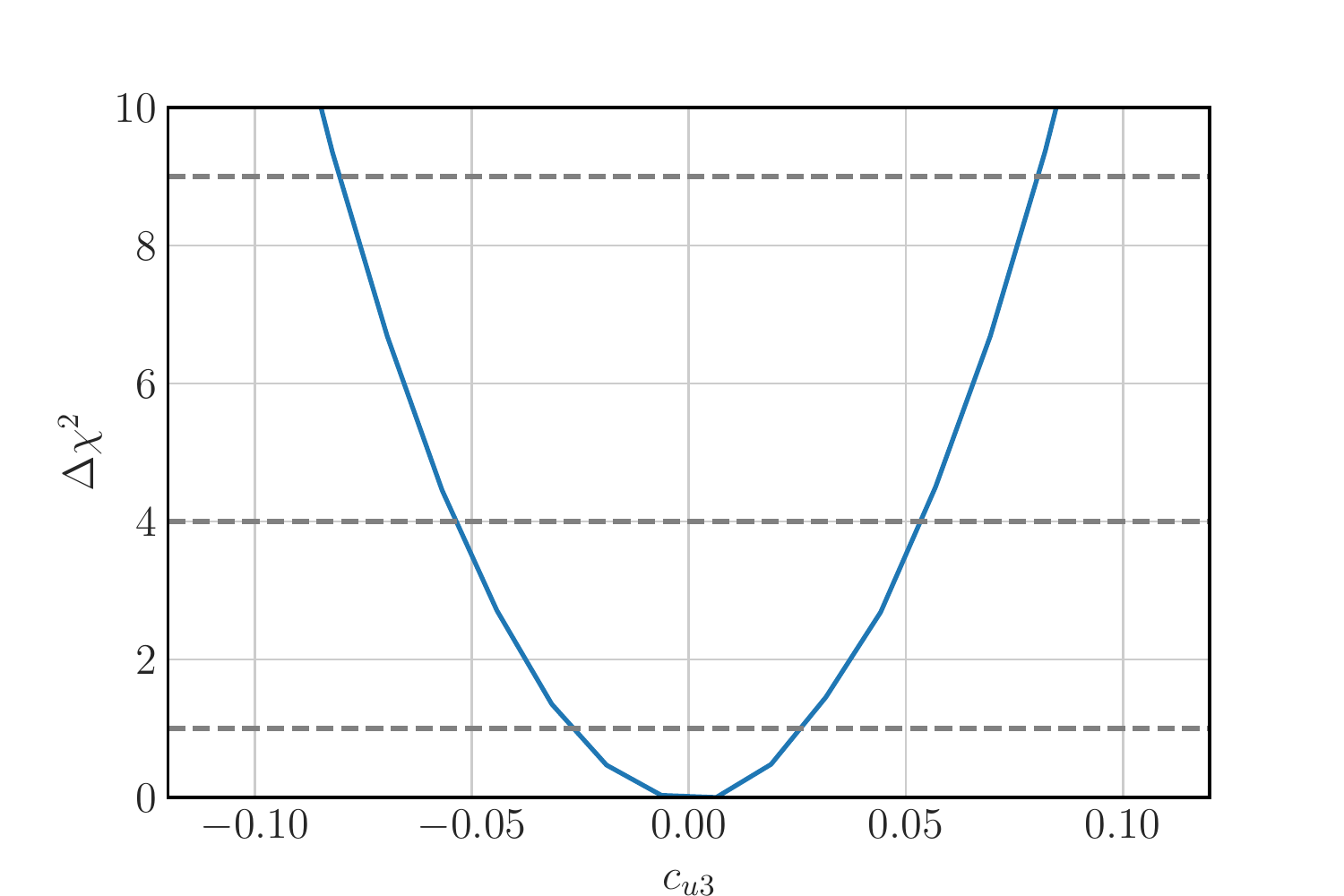}
  \includegraphics[width=0.45\textwidth]{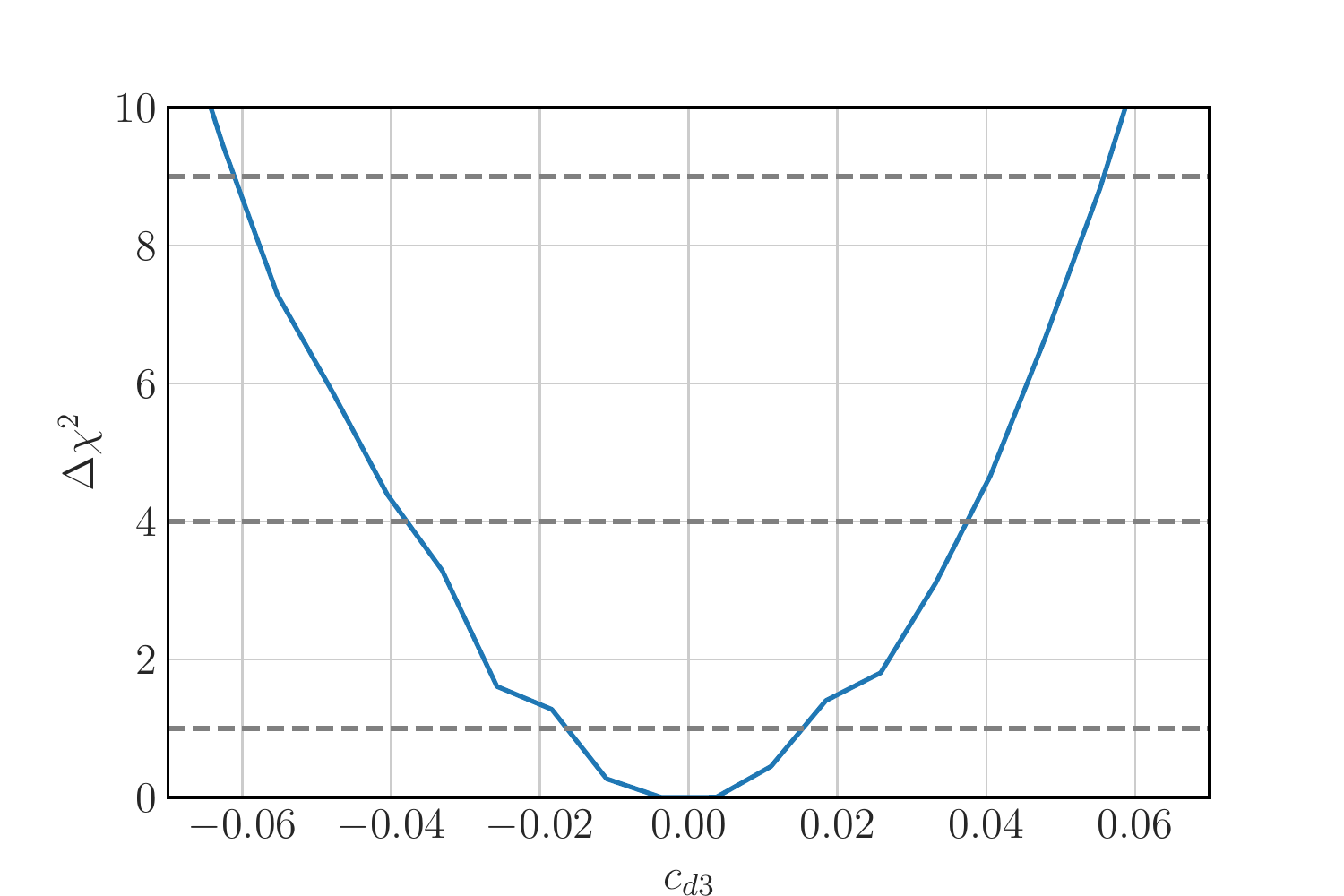}
  \includegraphics[width=0.45\textwidth]{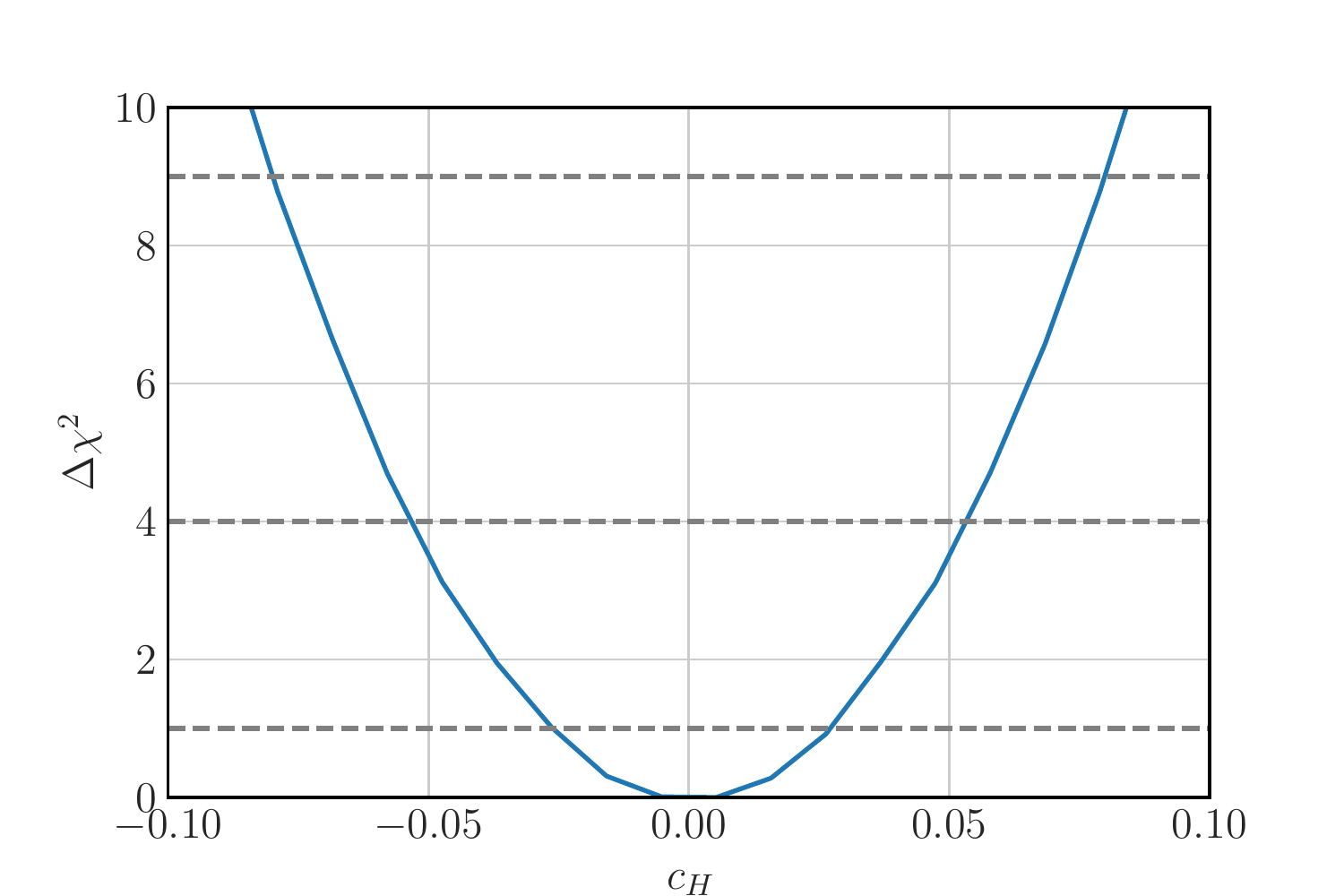}
  \includegraphics[width=0.45\textwidth]{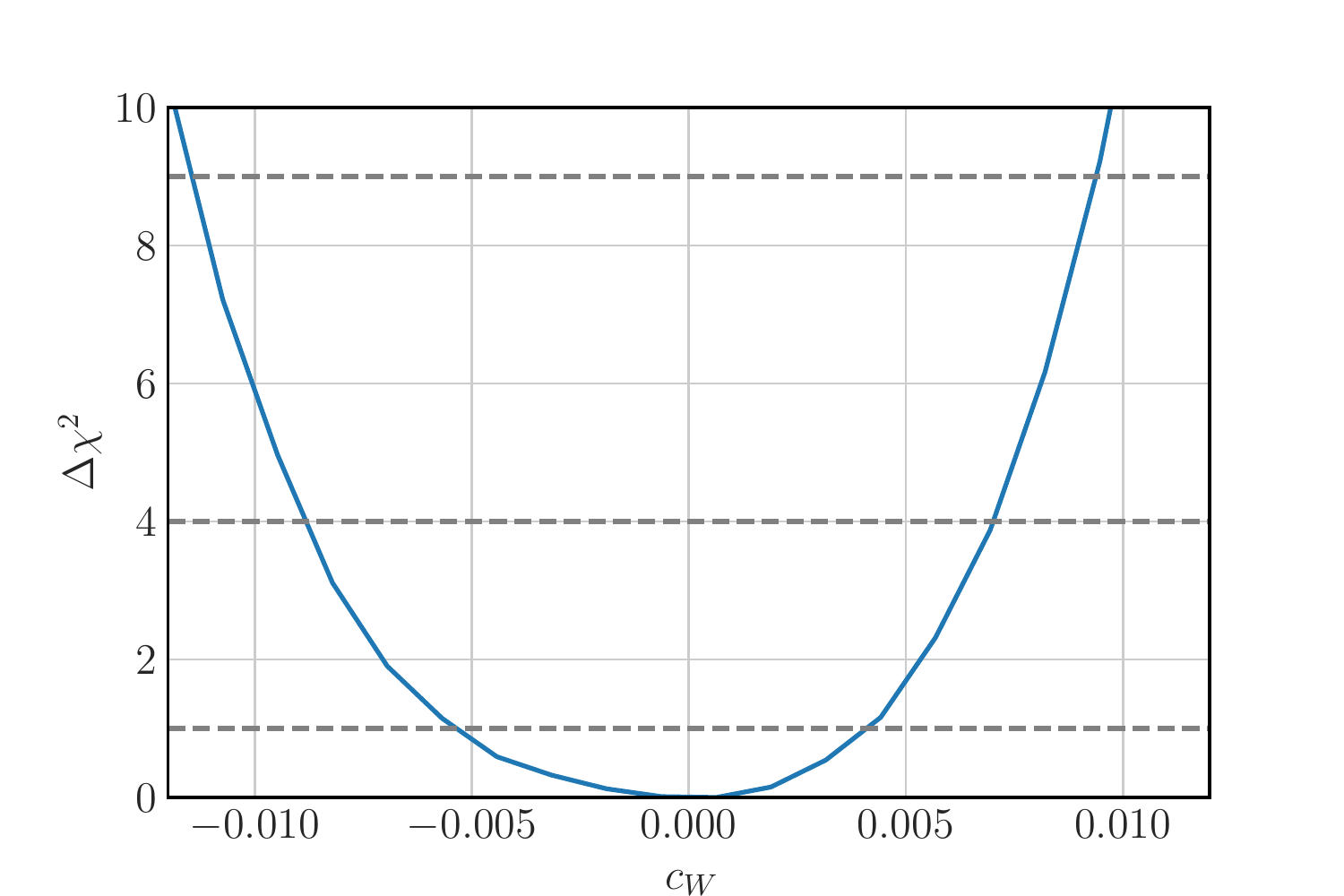}
  \includegraphics[width=0.45\textwidth]{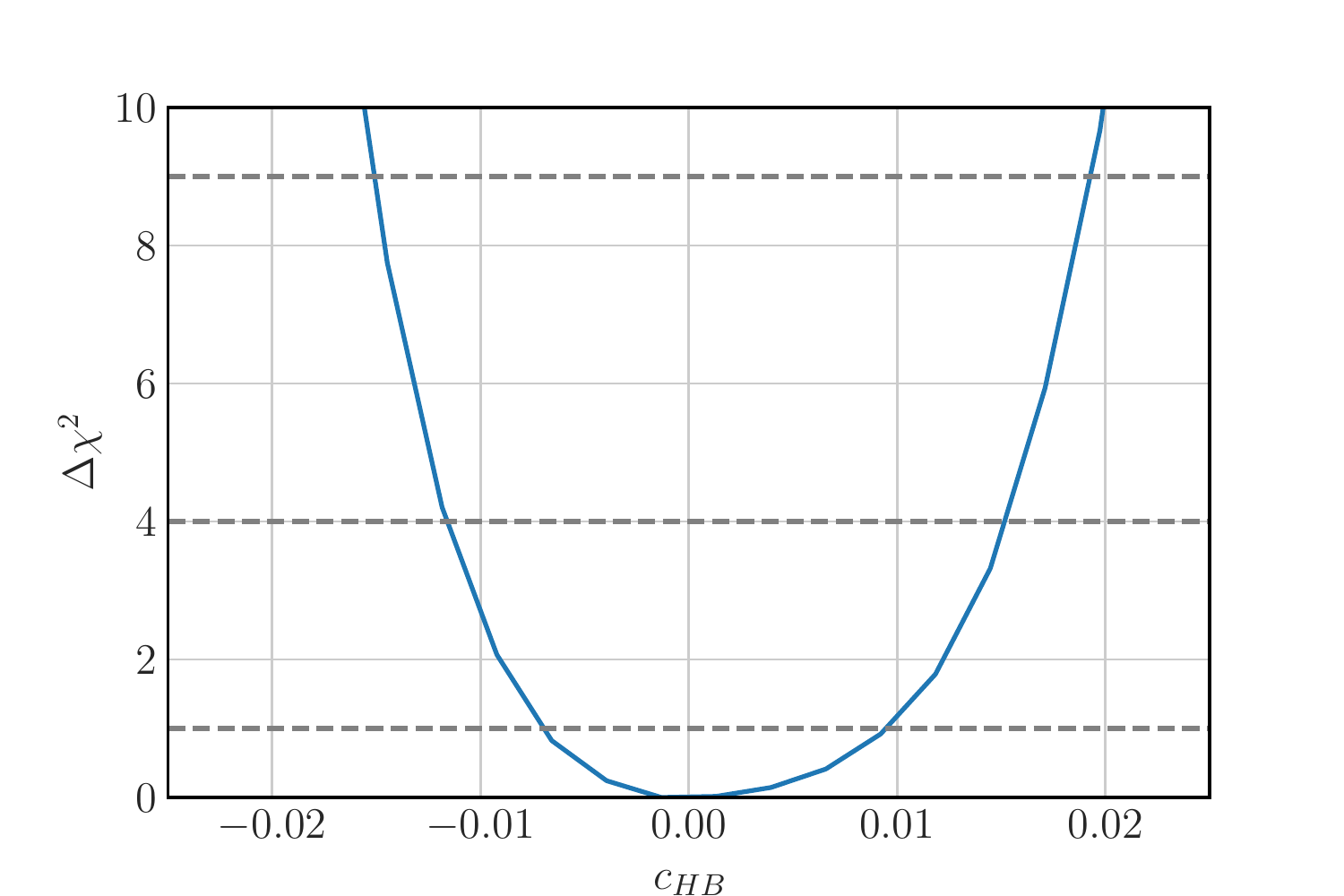}
  \includegraphics[width=0.45\textwidth]{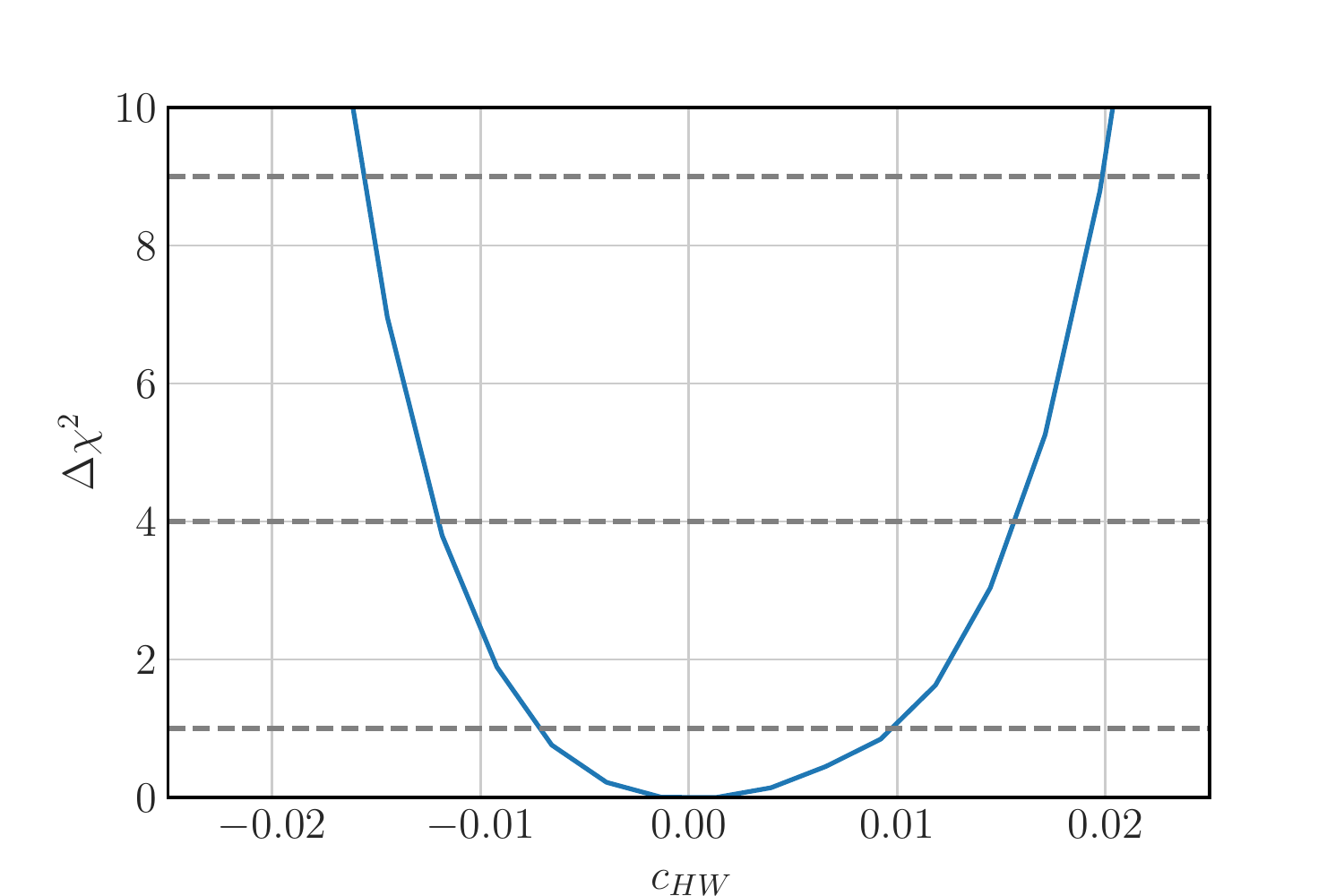}
  \caption{Minimal $\Delta \chi^2 = \chi^2 - \chi^2_{\text{min}}$ as a function of the each of the Wilson coefficients in the Higgs fit. }
  \label{fig:higgs-profiles}
\end{figure*}

\begin{figure*}
  \centering
  \includegraphics[width=0.29\textwidth]{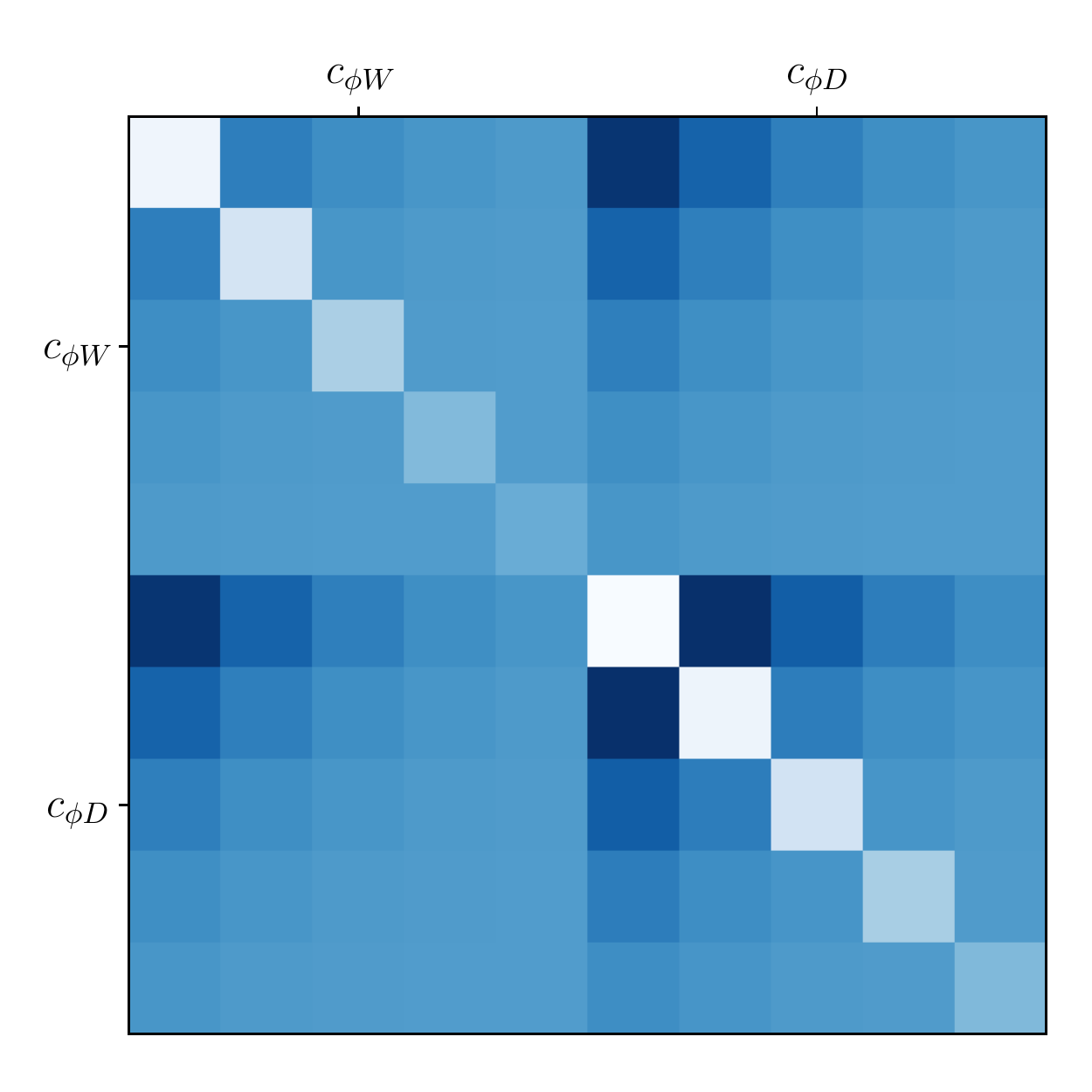}
  \includegraphics[width=0.308\textwidth]{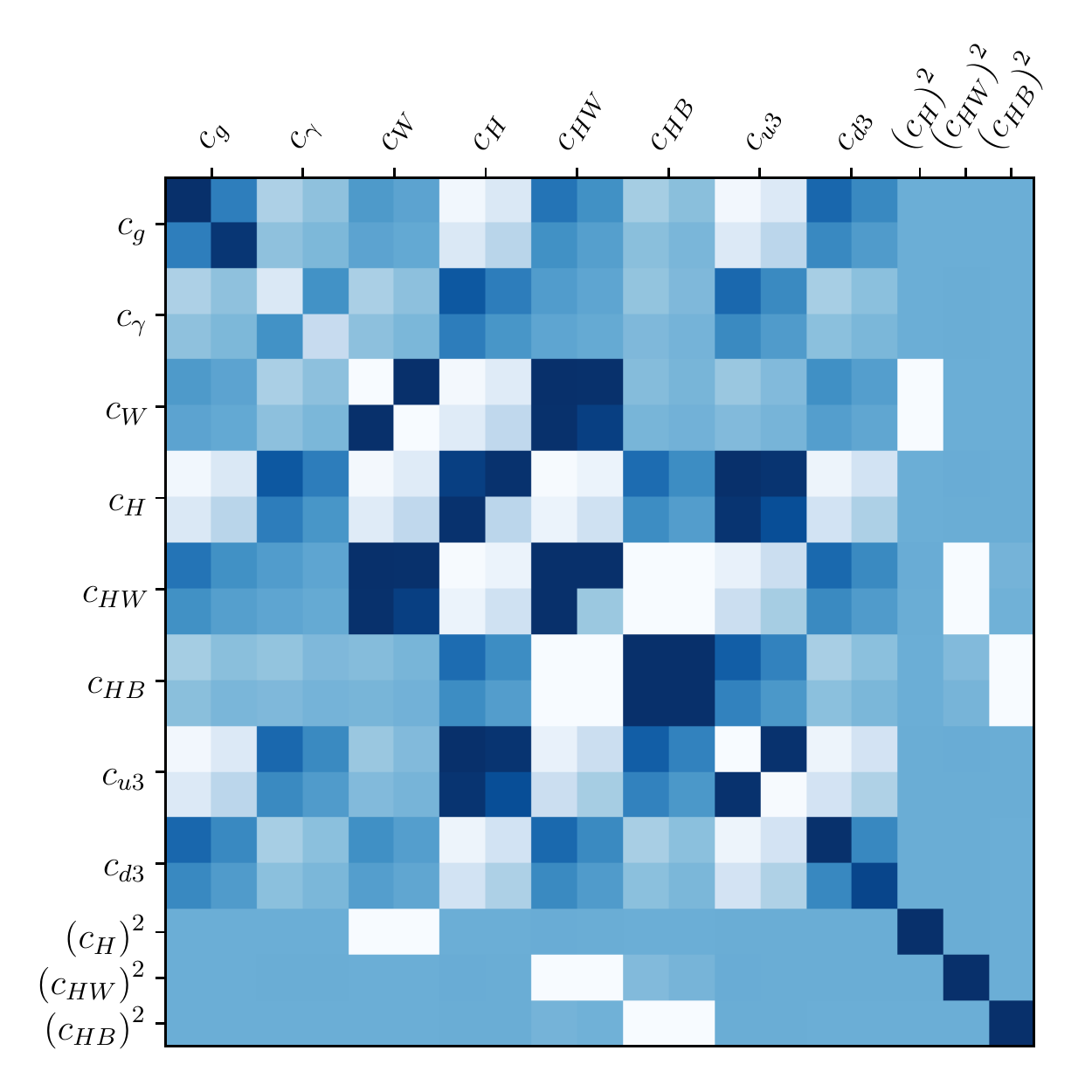}
  \includegraphics[width=0.29\textwidth]{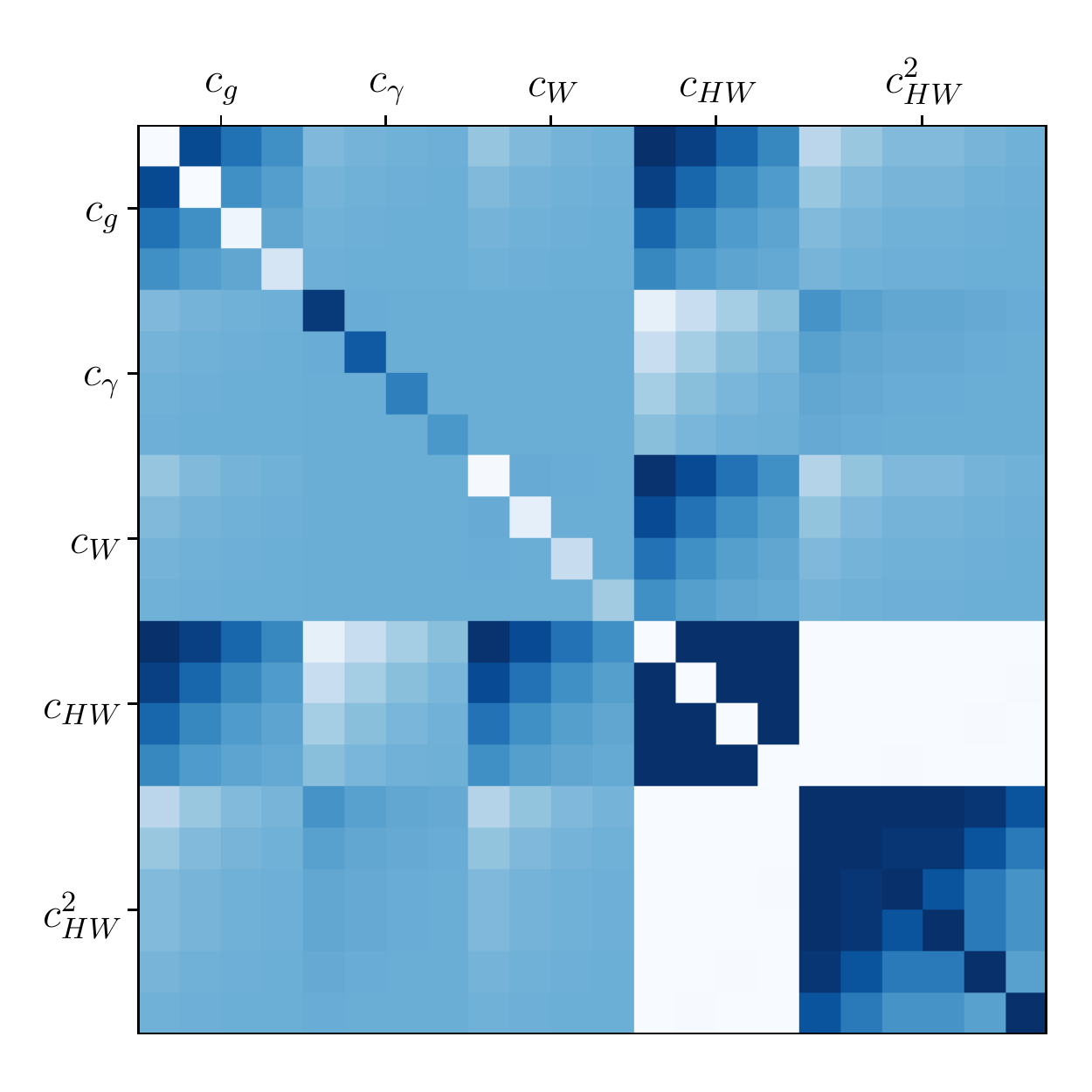}
  \includegraphics[width=0.08\textwidth]{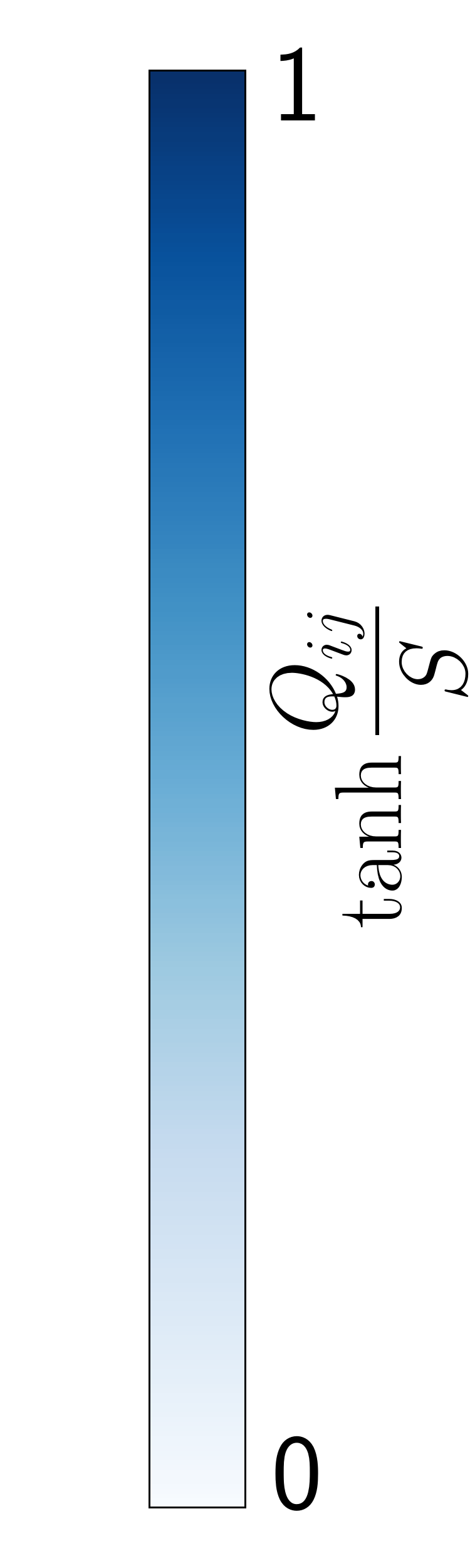}
  \caption{Qubo matrices $Q$ for the first zooming epoch of the EWPO (left), Higgs (center) and non-convex (right) fits. The color code shows the value of $\operatorname{tanh}(Q_{ij} / S)$, where $S = 10, 10^2, 10^4$ for the EWPO, Higgs and non-convex fits, respectively. The lowest value of the matrix is shown in white, while the highest is represented in dark blue.}
  \label{fig:qubos}
\end{figure*}

As a first and simple test of the method, we perform a fit of the $S$ and $T$ oblique parameters to the electroweak precision observables. We use the set of such observables defined Ref.~\cite{Aebischer:2018iyb}, with the values, uncertainties and correlations given in Refs.~\cite{ALEPH:2005ab, ALEPH:2013dgf, CDF:2013dpa, ParticleDataGroup:2016lqr, ATLAS:2017rzl}. The $S$ and $T$ parameters can be defined in this context as
\begin{equation}
 S = \frac{16 \pi v^2}{g g'} c_{\phi WB}, \quad
 T = \frac{2 \pi \left(g^2 + (g')^2\right) v^2}{g^2 (g')^2} c_{\phi D},
\end{equation}
where $v$ is the Higgs vacuum expectation value, $g$ and $g'$ are the gauge coupling constants for the $SU(2)$ and $U(1)$ factors of the electroweak symmetry, and $c_{\phi WB}$ and $c_{\phi D}$ are the Wilson coefficients for the following dimension-6 operators in the Standard Model EFT:
\begin{equation}
 \mathcal{O}_{\phi W B} = (\phi^\dagger \sigma^a \phi) W^a_{\mu\nu} B^{\mu\nu}, \quad
 \mathcal{O}_{\phi D} = |\phi^\dagger D_\mu \phi|^2.
\end{equation}
Here, $\phi$ is the Higgs doublet, $W^a_{\mu\nu}$ and $B_{\mu\nu}$ are the $SU(2)$ and $U(1)$ field-strength tensors, $D_\mu$ is the corresponding covariant derivative, and $\sigma^a$ are the Pauli matrices.

We choose the following hyperparameters for the fit: $N = 5$ binary variables per coefficient, ranges for them given by $c_{\phi WB} \in [-0.005,0.005] \, \si{TeV^{-2}}$ and $c_{\phi D} \in [-0.03, 0.03] \, \si{TeV^{-2}}$, no zooming, 100 reads, and a linear anneal schedule $s(t)$, with annealing time $t_f - t_i = \SI{50}{\mu s}$. A visual representation of the $Q$ matrix of the corresponding QUBO problem is shown in Fig.~\ref{fig:qubos}. We implement this in the D-Wave \texttt{Advantage\_system4.1} quantum annealer. The total computing time spent in the quantum annealer for the fit is thus $\SI{5}{ms}$. Since there is no zooming procedure involved, the fit is performed purely quantum mechanically in this case, with no classical updating steps. As the result, we obtain the central values $S = 0.05$ and $T = 0.09$, close to the values obtained in similar fits in the literature~\cite{Haller:2018nnx}.

As the next step, we generate the profile of the minimal $\Delta \chi^2 = \chi^2 - \chi^2_{\text{min}}$ as a function of the parameter $S$. To do so, we perform the fit while keeping the value of $S$ fixed. The ranges of for the coefficients are taken in this case to be $c_{\phi WB} \in [-0.03, 0.03] \, \si{TeV^{-2}}$ and $c_{\phi D} \in [-0.1, 0.04] \, \si{TeV^{-2}}$, while the rest of hyperparameters are kept the same as in the fit for the central values. We repeat this for 20 different values of $S$, equally spaced in a range around its central value. We use the same procedure to produce the profile for $\Delta \chi^2$ as a function of $T$. The results are shown in Fig.~\ref{fig:ewpo-profiles}. From these profiles, we can derive the following $1\sigma$-intervals for the parameters:
\begin{equation}
 S = 0.05 \pm 0.09, \qquad T = 0.09 \pm 0.08,
\end{equation}
which are again close to those obtained in Ref.~\cite{Haller:2018nnx}. The total quantum annealing time employed in generating the profiles is $\SI{0.2}{s}$.

\begin{table*}
 \centering
 \begin{tabular}{cccccccc}
  \toprule
  Formulation
  & Method & Fit time
  & $c_{HW}$ & $c_H$ & $c_g$ & $c_\gamma$ & $\chi^2$\\
  \midrule
  \multirow{4}{*}{Standard}
  & Minuit (initial $c_{HW} = 0$) & $\SI{2.0}{s}$ 
  & -0.009 & 0.100 & \num{1.4e-5} & \num{3.2e-6} & 4110
  \\
  & Minuit (initial $c_{HW} = -0.05$) & $\SI{2.4}{s}$
  & -0.050 & 0.039 & \num{-9.7e-6} & \num{-1.0e-4} & 135 \\
  & Simulated annealing (initial $c_{HW} = 0$) & $\SI{642}{s}$
  & -0.009 & 0.100 & \num{1.4e-5} & \num{3.7e-6} & 4110 \\
  & Simulated annealing (initial $c_{HW} = -0.05$) & $\SI{644}{s}$
  & -0.009 & 0.100 & \num{1.4e-5} & \num{3.7e-6} & 4110 \\   
  \midrule
  \multirow{3}{*}{QUBO} & Simulated annealing (Class A) & $\SI{6.4}{s}$
  & -0.012 & -0.054 & \num{-3.0e-5} & \num{3.9e-5} & 3910 \\
  & Simulated annealing (Class B) & $\SI{6.4}{s}$
  & -0.045 & -0.175 & \num{-3.7e-5} & \num{1.8e-4} & 228 \\
  & Quantum annealing & $\SI{0.2}{s}$
  & -0.047 & -0.050 & \num{1.9e-5} & \num{7.5e-7} & 68 \\
  \bottomrule
 \end{tabular}
 \caption{Best fit values of the Wilson coefficients from different fitting methods, for the non-convex $\chi^2$ Higgs fit. The QUBO-formulation simulated annealing generates radically different results for different runs with the same parameters, which fall into 3 classes: (A) $\chi^2 \simeq 4000$, (B) $\chi^2 < 1000$ and (C) $\chi^2 > 10^7$, occurring $13\%$, $82\%$ and $5\%$ of the time, respectively. We present here an example of class-A and an example class-B results.}
 \label{tab:nonconvex}
\end{table*}

\subsection{Higgs fit}
\label{sec:higgs-fit}

We now use our method to perform a larger fit, including 8 Wilson coefficients and quadratic terms in the theoretical predictions for observables. The Lagrangian we consider is a subset of the dimension-6 Standard Model EFT Lagrangian, given by:
\begin{align}
 \mathcal{L}
 &=
	\frac{c_{u3} y_t}{v^2} (\phi^\dagger \phi) (\bar{q}_L \tilde{\phi} u_R)
	+ \frac{c_{d3} y_b}{v^2} (\phi^\dagger \phi) (\bar{q}_L \phi d_R)
 \nonumber \\
 &
	+ \frac{ic_W g}{2 m^2_W} (\phi^\dagger \sigma^a D^\mu \phi) D^\nu W^a_{\mu\nu}
	+ \frac{c_H}{4 v^2} \left(\partial_\mu (\phi^\dagger \phi)\right)^2
 \nonumber \\
 &
	+ \frac{c_\gamma (g')^2}{2 m^2_W} (\phi^\dagger \phi) B_{\mu\nu} B^{\mu\nu}
	+ \frac{c_g g_S^2}{2 m^2_W} (\phi^\dagger \phi) G^a_{\mu\nu} G^{a\mu\nu}
 \nonumber \\
 &
	+ \frac{ic_{HW} g}{4 m^2_W} (\phi^\dagger \sigma^a D^\mu \phi) D^\nu W^a_{\mu\nu}
 \nonumber \\
 &
	+ \frac{ic_{HB} g'}{4 m^2_W} (\phi^\dagger D^\mu \phi) D^\nu B_{\mu\nu}
	+ \text{h.c.}
\end{align}
where $y_t$ and $y_b$ are the top and bottom Yukawa couplings, $m_W$ is the mass of the $W$ boson, and $g_S$ is the strong coupling constant.

We use the set of observables described in Ref.~\cite{Englert:2017aqb}. They correspond to projections of Higgs production and decay processes at the High-Luminosity LHC. We consider only inclusive measurements in gluon fusion, vector boson fusion, and the production associated with an electroweak gauge boson, two top quarks or a jet. The decays are into the following final states: $\gamma\gamma$, $W^+W^-$, $ZZ$, $\mu^+\mu^-$, $\tau^+\tau^-$ and $b\bar{b}$.

We compute the theoretical predictions for the observables, up to quadratic order in the coefficients. We find that $c_H$, $c_W$, $c_{HB}$ and $c_{HW}$ have quadratic contributions to some observables. We only consider contributions of the form $c_i^2$ for observables where a linear parametrization results in a poor description of the parameter dependence\footnote{
We parametrize the $c_i$ dependence of each observable by a fit to points generated on a grid and demand $\chi^2<0.03$ for a linear parametrization. If $\chi^2>0.03$, we use a quadratic parametrization in $c_i$. 
}. The only coefficients that have non-negligible contributions at the quadratic level are $c_W$, $c_{HB}$ and $c_{HW}$. We assume that the measured values of the coefficients are the same as their (dimension-4) Standard Model values. We use the QUBO formulation of the fit to generate the individual $\Delta \chi^2$ profiles for the coefficients.

There are several choices of hyperparameters that give rise to similar results. However, we find a greater consistency among different trial runs with the following set of values: $N=2$ binary variables per coefficient, 20 zooming epochs, a zoom factor of $f = 80\%$, 200 reads in each epoch, and $t_f - t_i = \SI{100}{\mu s}$ of annealing time for each read, with a linear anneal schedule $s(t) \propto t - t_i$. The ranges we use for the coefficients are centred around the origin ($L_i = -U_i$), with the following upper limits:
\begin{multline}
 U_{u3} = 2U_{d3} = 10^4 U_g = 10^5 U_\gamma = \\
 = U_H = 10 U_W = 5 U_{HB} = 5 U_{HW} = 0.1,
\end{multline}
The matrix $Q$ for the corresponding QUBO problem is shown in Fig.~\ref{fig:qubos}. Again, we use the D-Wave \texttt{Advantage\_system4.1} quantum annealer to perform the fits. The total quantum annealing time per fit is $\SI{0.4}{s}$.

We show our results in Fig.~\ref{fig:higgs-profiles}. For the coefficients $c_g$, $c_\gamma$, $c_H$, $c_{u3}$ and $c_{d3}$, we obtain a $\Delta \chi^2$ profile with quadratic shape, which is symmetric under $c_i \to -c_i$, as only have linear contributions to the observables. For $c_W$, $c_{HB}$ and $c_{HW}$, we observe an asymmetry under $c_i \to -c_i$, generated by the corresponding quadratic terms.

\subsection{Non-convex Higgs fit}
\label{sec:mod-higgs-fit}

Finally, we consider a modified scenario in which the $\chi^2$ function has a local minimum close to $c_i = 0$ and a (much lower) global one away from it. We thus have to minimize non-convex loss functions, a problem in which the quantum approach might provide an advantage over its classical counterparts~\cite{Abel:2021fpn}.

To generate this setup, we consider a scenario in which several observables have been measured to be away from their Standard Model values. The observable that plays the most important role in generating the non-convexity is $\mu_{pp \to WH \to WZ\gamma}$, which we set to $-0.3$, with total uncertainty to be $\Delta \mu_{pp \to WH \to WZ\gamma} = 0.01$. For the rest of the introduced deviations from the Standard Model values, we keep the projected experimental uncertainties while setting the measured values as follows:
\begin{itemize}
 \item For gluon fusion, vector-boson function, $ttH$ and $jH$ processes with $WW$ or $ZZ$ in the final state, we take $\mu_i = 0.8$.
 \item For any process with $Z\gamma$ is in the final state, we set $\mu_i = -6$.
 \item For all the processes with Higgs production in association with a vector boson, we choose $\mu_i = 0.2$.
\end{itemize}
We keep the remaining observables in the fit at their Standard Model values. Concerning the coefficients, we only allow $c_{HW}$, $c_W$, $c_g$ and $c_\gamma$ to vary, while fixing the rest of the coefficients to zero.

For the fits in this section, we use the QUBO formulation with $N = 4$ binary variables per coefficient. The hyperparameters are chosen the same as in Section~\ref{sec:higgs-fit}: 20 zooming epochs, a zooming factor $f = 80\%$, 200 reads per epoch, and $t_f - t_i = \SI{100}{\mu s}$ annealing time, with a linear schedule $s(t)$. The ranges of values for the coefficients that we choose are $c_g, c_\gamma \in [-2, 2] \times 10^{-4}$, $c_{HW} \in [-0.1, 0]$, $c_W \in [-0.1, 0.1]$. The $Q$ for the QUBO problem corresponding to the first zooming epoch is shown in Fig.~\ref{fig:qubos}.

\begin{figure}
 \centering
 \includegraphics[width=0.45\textwidth]{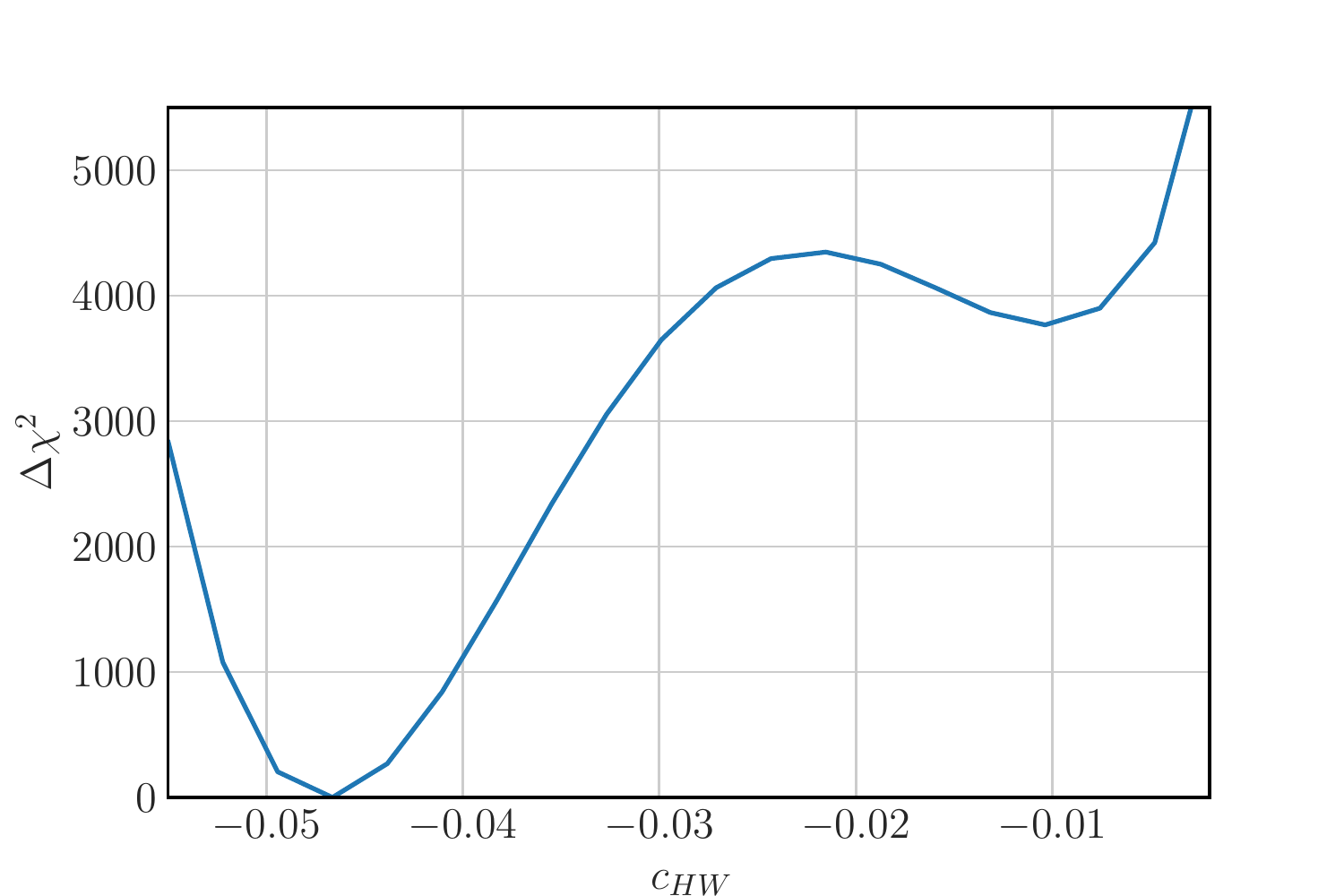}
 \caption{$\Delta \chi^2 = \chi^2 - \chi^2_{\text{min}}$ as a function of $c_{HW}$, displaying a local minimum close to $c_{HW} = 0$ and a global one near $c_{HW} = -0.05$.}
 \label{fig:nonconvex-profile}
\end{figure}

We embed this QUBO in the D-Wave \texttt{Advantage\_system4.1} quantum annealer, fixing $c_{HW}$ at 20 equally-spaced values in its range to generate the $\Delta \chi^2$ profile for it. The results are given in Fig.~\ref{fig:nonconvex-profile}. This suggests that $\chi^2$ indeed has 2 local minima, one at $c_{HW} \simeq -0.01$ and a lower one at $c_{HW} \simeq -0.05$. We check that the gradients are small at these two points and that the Hessian is definite positive. As a quantitative measure of the smallness of the gradients, we find that the following equation is satisfied:
\begin{equation}
 \sum_i \frac{c_i}{10} \frac{\partial_{c_i} \chi^2}{\chi^2} < 10 \%.
\end{equation}
We conclude that we have found an excellent approximation to two of the local minima of $\chi^2$. As shown in Appendix~\ref{sec:two-minima}, the form of the $\chi^2$ function that we consider here implies that we can have at most two local minima, so we have identified all of them. With $\chi^2 \geq 0$, this implies that the minimum at $c_{HW} \simeq -0.05$ is the global one.

We now compare different methods for finding the maximum-likelihood estimations of the four coefficients. This corresponds to a search for the global minimum of $\chi^2$. The classical methods that we consider are:
\begin{itemize}
 \item The \texttt{MIGRAD} algorithm provided by the \texttt{Minuit} code~\cite{James:1975dr}. This type of variable-metric algorithm works by incrementally improving an approximation of the error matrix and the best-fit point.
 \item Simulated annealing. This method also proceeds by iterativaly improving a point in parameter space. In each step, the algorithm moves to a random nearby point if the new point has a lower $\chi^2$. If the new point has a higher $\chi^2$, it moves to it with probability $e^{-\Delta \chi^2 / T}$, where $\Delta \chi^2$ is the difference in $\chi^2$ between the new and the old points, and $T$ is a parameter that decreases along the run, known as the temperature. The algorithm explores the vicinity of local minima while changing $T$ depending on the number of failures to find a better minimum in the previous steps. This may allow the algorithm to escape local minima in some cases. We chose the initial value of $T$ to be $10^6$ with a minimum value of $10^{-6}$, an adaptive speed of $1$ and a step size of $0.01$. The maximum number of steps is chosen to be $10^5$.
 \item Finally, we also consider the same algorithm as for the quantum annealing approach, with the QUBO formulation and a zooming process, replacing the quantum annealing runs with simulated annealing. In this case, the parameter points to be updated by the algorithm are the sets of $0$ or $1$ values of the binary variables. The random nearby point is obtained by flipping a random variable. The number of steps is usually measured here in terms of \emph{sweeps}, which correspond to as many updating steps as binary variables are present in the problem.
\end{itemize}
We refer to the methods not using the QUBO formulation as \emph{standard-formulation} methods.
We show a comparison of the best-fit values of the coefficients and $\chi^2$ obtained from all the methods in Table~\ref{tab:nonconvex}. The classical methods are initialized to a point with $c_W = c_g = c_\gamma = 0$.

Inspecting the $\chi^2$ values given in the Table~\ref{tab:nonconvex}, one can see that classical algorithms that start at the point $c_i = 0$ typically get trapped in the local minimum nearby. This problem is not present in the quantum annealing approach. The minimum with $\chi^2 = 135$ obtained by Minuit for initial point $c_{HW} = -0.05$ corresponds to the one from the quantum annealer. The small differences in the minimum value of $\chi^2$ and the best-fit parameters are because we have neglected some quadratic contributions in the formulation of the QUBO problem, which has no consequences on the general shape of the $\chi^2$ function, with a global minimum around $c_{HW} = -0.05$ and a local one close to $c_{HW} = 0$.

The QUBO-formulation simulated annealing results depend strongly on the schedule for the temperature. For large starting temperatures, there is no dependence on the starting point. The results also vary considerably for fixed annealing parameters from run to run. We find the most consistent results with a schedule that exponentially increases the inverse-temperature parameter $\beta$ from $\num{1e-5}$ to $10$ in $\num{1e6}$ steps, performing one sweep per step. The results fall into three classes: (A) those on the ``wrong'' side of the barrier, with $\chi^2 \simeq 4000$; (B) those on the correct side, with $\chi^2 < 1000$; and (C) those for which the zooming gets stuck at $\chi^2 > 10^7$. We perform 40 runs and find that $13\%$ of the runs belong to class A, $82\%$ to class B, and $5\%$ to class C. The results in class (B) also present a considerable variation. We show an arbitrary example in Table~\ref{tab:nonconvex}.

In Table~\ref{tab:nonconvex}, we also provide the total time spent performing the optimization for each method. Again, we run Minuit, the standard-formulation simulated annealing and the QUBO simulated annealing in an Apple M1 processor. Since quantum annealing is performed on a dedicated device, the numbers cannot be compared directly. However, we note that quantum annealing requires orders of magnitude less time to perform this task.

\section{Conclusions}
\label{sec:conclusions}

We have presented QFitter, a quantum annealing-based method for fitting EFT coefficients to experimental measurements. The $\chi^2$ is encoded as a QUBO problem which can be directly embedded in the currently-available quantum annealers. The required number of qubits depends on the number of coefficients to be fitted, the non-linear terms in their contributions to observables (which require auxiliary qubits), and the precision to which they are to be determined. The number of observables included does not affect this.

Since physical annealers only provide a limited amount of qubits, the practical implementation of QFitter can only be done for a limited number of coefficients, precision and non-linearities. We have used a zooming algorithm, in which the precision is increased iteratively through several annealing runs, to overcome this limitation partially. With this setup, we have found that fitting problems involving at least eight coefficients and their quadratic dependencies can be embedded in current quantum annealing devices.

Finally, we have tested the performance of QFitter with three examples. The first two, the EWPO and the Higgs fit, involve a convex $\chi^2$ function. The quantum approach gives comparable results to the classical ones here. We have then modified the $\chi^2$ for the Higgs fit to make it non-convex. By comparing with several classical algorithms, we have found that the quantum one is the one that ends in the global minimum most consistently with a considerable gain in processing time. 

\vspace{10pt}

\appendix

\section{Number of local minima in the non-convex Higgs fit}
\label{sec:two-minima}

We prove here that the $\chi^2$ function defined in Section~\ref{sec:mod-higgs-fit} has 2 local minima at the most. Neglecting constant terms, $\chi^2$ can be written as
\begin{equation}
  \chi^2 =  R_{ij} x_i x_j + S_{ij} y_i y_j + T_{ij} x_i y_j,
\end{equation}
where the $R_{ij} = R_{ji}$, $S_{ij} = S_{ji}$ and $T_{ij}$ are constant parameters, and $x = (c_W, c_g, c_\gamma)$, $y = (c_{HW}, c_{HW}^2)$. It can be checked that the matrix $R$ is invertible. At a local minimum we must have $\nabla \chi^2 = 0$. In particular, $\partial_{x_i} \chi^2 = 0$ must be satisfied. That is,
\begin{equation}
  2 R_{ij} x_j = -T_{ij} y_j.
  \label{eq:app-necessary}
\end{equation}
This is a system of linear equations in $x_j$ with a unique solution $\bar{x}_i(y) = -\frac{1}{2} R_{ik}^{-1} T_{kj} y_j$. Since Eq.~\eqref{eq:app-necessary} is a necessary condition for a point $(x, y)$ to be a local minimum, all the local minima of $\chi^2$ must lie on the curve described by $(\bar{x}(y), y)$. We can thus look for the local minima of
\begin{equation}
  \chi^2|_{x_i = \bar{x}_i(y)}
  =
  R_{ij} \bar{x}_i(y) \bar{x}_j(y) + S_{ij} y_i y_j + T_{ij} \bar{x}_i(y) y_j.
\end{equation}
The fact that $\bar{x}_i(y)$ is linear in $y$ implies that it is quadratic in $c_{HW}$. It follows that $\chi^2|_{x_i = \bar{x}_i(y)}$ is a quartic polynomial in $c_{HW}$. This concludes the proof, since a quartic polynomial can only have up to 2 local minima.

\bibliography{references}

\end{document}